\documentclass[sigconf,nonacm]{acmart}

\renewcommand\footnotetextcopyrightpermission[1]{}
\acmConference[AITC 2026]{AI Transparency Conference}{June 5--6, 2026}{TBD}

\usepackage{subcaption}    

\usepackage{booktabs}
\usepackage{multirow}

\usepackage[most]{tcolorbox}
\usepackage{xcolor}

\usepackage{enumitem}

\newtcolorbox{promptbox}{
  colback=gray!15,
  colframe=gray!15,
  boxrule=0pt,
  arc=0pt,
  left=6pt,
  right=6pt,
  top=4pt,
  bottom=4pt,
  boxsep=0pt,
  fontupper=\ttfamily,
  before skip=4pt,
  after skip=8pt
}

\usepackage{todonotes}

\begin{document}

\title{The Energy Society: A Simulation Environment for Studying Agent Cooperation under Survival Pressure}

\author{Lucas Bergholdt Hansen}
\affiliation{
    \institution{University of Southern Denmark}
    \city{Odense}
    \country{Denmark}
}
\email{lucha23@student.sdu.dk}

\author{Federico Torrielli}
\orcid{0000-0001-8037-8828}
\affiliation{
    \institution{University of Turin}
    \city{Torino}
    \country{Italy}
}
\email{federico.torrielli@unito.it}

\author{Filippo Tonini}
\orcid{0009-0001-4288-9384}
\affiliation{
  \institution{University of Southern Denmark}
  \city{Odense}
  \country{Denmark}
}
\email{tonini@imada.sdu.dk}

\author{Lukas Galke Poech}
\orcid{0000-0001-6124-1092}
\affiliation{
  \institution{University of Southern Denmark}
  \city{Odense}
  \country{Denmark}
}
\email{galke@imada.sdu.dk}

\begin{abstract}
LLM-based agents are increasingly deployed in multi-agent environments whose incentives can shape their behavior. We introduce The Energy Society, a minimal survival economy for studying how competitive and cooperative incentives affect emergent behavior when inference cost is directly tied to survival: Agents spend energy based on model size when generating tokens, regain energy by completing jobs or receiving donations, and deactivate if their energy reaches zero.
We compare competitive and cooperative objectives against a baseline setting and several control variants.
Across experiments, larger models consistently consume the most energy and spend more energy than they gain, even in those settings where token cost is not size-dependent.
Cooperative incentives substantially alter behavior: agents donate to reactivate others, sometimes at the cost of their own survival, and job allocation changes. 
Ablations reveal that allowing agents to recommend actions to each other supports coordination and ambitious job selection, while memory helps agents calibrate risk from past outcomes.
Agents rarely choose direct sabotage, but show more subtle signs of self-serving behavior in the competitive setting.
The Energy Society is a compact testbed for studying the interaction between  token costs and group incentives under a survival pressure. Source code is available at: \url{https://github.com/LucasBergholdt/EnergySociety}
\end{abstract}

\keywords{Multi-Agent Systems, LLM Agents, Emergent Cooperation, Simulation Environment, Survival Pressure, Resource-Constrained Agents}

\maketitle

\section{Introduction}
Large Language Models have recently risen in popularity as the basis for autonomous agents. 
Agents that observe an environment, maintain state, interact with other agents, make decisions, and adapt behavior over repeated interactions. This shift has renewed interest in the long-standing idea of computationally-powered agents that react in an appropriate manner to their environment.
The vision is that such agents could be used to modeling social phenomena \cite{10.1145/3526113.3545616, Dill2011AGA}, be employed in training systems \cite{10.1609/aimag.v5i2.434, 10.1609/aimag.v16i1.1121, 10.5555/648034.744239}, test social science theories \cite{horton2026largelanguagemodelssimulated, Binz_2023}, power robots \cite{10.1145/176789.176803}, and enhance non-playable characters to create more immersive gaming experiences \cite{10.5555/2900929.2901035, 10.1609/aimag.v22i2.1558}.

While earlier agents often relied on hand-crafted rules, fixed policies, or task-specific architectures, LLMs may offer a more general-purpose underlying intelligence.
Recent work has shown that LLMs deployed as agentic systems within multi-agent environments can exhibit emergent behaviors and social dynamics \cite{chen2023agentversefacilitatingmultiagentcollaboration}. This has motivated simulation-based studies across a range of settings, from sandboxes displaying human-like behavior \cite{park2023generativeagentsinteractivesimulacra}, to simulations of communication within and between alien civilizations \cite{xue2025llmsdifferentworldviews}. 
However, one important aspect of LLM-based agents is often left implicit: the cost of inference. In many simulations, agents may reason extensively and produce long outputs without the cost of generating those tokens directly affecting the simulation. This abstraction is useful, but differs both from real-world deployments, where token generation has a cost. This constraint is to some extent analogous to human decision-making: cognitive effort is not free. 

In this paper, we introduce the Energy Society, a minimal multi-agent simulation environment in which an agent's token usage is directly linked to its survival\footnote{We use the term survival analogically: agents that exhaust their token budget cease to operate, mirroring selection under resource scarcity, without implying that LLM agents are living organisms.}:
Each day, agents recommend actions to one another, select an action independently, and execute their chosen action.
Each agent begins with a finite energy reserve, spends energy whenever they generate tokens scaled by the size of their model, regains energy by completing jobs or receiving donations, and deactivates if their energy reaches zero. 
This way, reasoning (i.e., generating tokens to form a chain of thought) is coupled with its computational cost, creating both a survival and efficiency pressure, and thereby incentivizing agents to estimate in advance not only what jobs they can complete but also whether the job is worth the cost.
We use the Energy Society to investigate to what extent LLM-based agents adapt their behavior when operating under individual vs. shared objectives in an environment that rewards coordination.
We compare two variants of the same environment: a competitive and cooperative condition, which differ in the instructed objective: maximizing own energy or maximizing the group's energy. 
To systematically isolate the factors driving agent behavior, we evaluate these two conditions across a baseline experiment and a series of subsequent controlled variants, each modifying a single environmental mechanic. We run all experiments across the same five seeds to ensure consistent job sampling, and analyze results using aggregate metrics.

Our results provide an initial empirical characterization of how LLM-based agents behave when reasoning, action and survival are tied to the same limited resource.
We find several recurring patterns:
Model size creates a persistent asymmetry in efficiency. Larger models consistently consume more energy and spend more than they gain, even when the size-dependent token cost is removed.
Shifting from the competitive to the cooperative setting changes job allocations and causes agents to start donating to reactivate other agents, sometimes at the cost of deactivating themselves.
Recommendations support coordination and agents use memory of previous rounds to calibrate job selection. Smaller agents show unexpected resilience under resource scarcity and an analysis of agent recommendations reveals they are sometimes used in a self-serving manner in the competitive setting.

In sum, our contributions are as follows:
\begin{itemize}
    \item We introduce The Energy Society, a multi-agent testbed for studying LLM agents under survival pressure linked to token generation, capable of showing emergent behaviors.
    \item We provide an empirical characterization of how cooperative and competitive incentives and the various environment mechanics affect agent behavior.
    \item We offer a discussion analyzing recurring patterns of observed behavior, which may be treated as more general hypotheses for future research.
\end{itemize}

\section{Related work}
\subsection{LLM-Based Agents in Simulated Environments}
The surge in large language model capabilities has sparked a new wave of research into autonomous AI agents and simulations of varied complex scenarios \cite{Wang2024, gao2023largelanguagemodelsempowered}. Because LLMs encode a wide range of human behavior from their training data, they can serve as simulacra of human-like behavior \cite{horton2026largelanguagemodelssimulated}.
LLM-based agents can interpret natural language observations, maintain histories of previous events, communicate with other agents, and select actions in response to changing environmental conditions.
If prompted with specific contextual descriptions, these agents can be used to investigate emergent behavior and social dynamics.
Recent applications of this approach have demonstrated its potency with simulations in which the behavior of individual agents and the dynamics between them emerge from repeated interactions with a structured environment.
Generative Agents \cite{park2023generativeagentsinteractivesimulacra} is a central example, introducing agents that populate an interactive sandbox world and produce believable behavior over time, with both individual routines and social interactions. A key part of their contribution is an agent architecture that stores experiences in natural language, retrieves relevant memories, synthesizes higher-level reflections, and uses these memories and reflections to determine actions and longer-term plans.
This architecture is well suited to their open-ended setting, however, \citet{dai2024artificialleviathanexploringsocial} demonstrates that comparatively simple prompt-level memory mechanisms (action logs) suffice to generate rich social dynamics in simpler settings.
The Energy Society adopts this simpler approach.

\subsection{Cooperation, Competition and Survival Pressure in LLM-based Multi-Agent Systems}

In multi-agent settings, agents may need to coordinate, divide tasks, compete for rewards, or decide whether to prioritize individual or collective outcomes.
\citet{tonini2025super} offer evidence for how repeated interactions and -- counterintuitively -- inter-group competition are essential for maintaining cooperative behavior in language model agents. This emphasizes the importance of interaction structures in enabling cooperation and shows that competition does not necessarily undermine cooperation.

Other work suggests that competitive survival pressure can also have negative effects. \citet{ma2025hungergamedebateemergence} found that introducing competitive survival pressure in a debate environment increased over-competition and degraded task performance, suggesting that competition under survival pressure can distort agent behavior when the environment rewards winning over accuracy or collaborative task performance.
By contrast, \citet{dai2024artificialleviathanexploringsocial} demonstrates that survival incentives and resource pressure can also lead to more cooperative social structures in a society simulation.
Together, these findings suggest that agent behavior is shaped not only by the model but also by the incentives and pressures imposed by the environment. While LLMs are typically cooperative by default, this behavior may change based on environmental factors \cite{fontana2024nicerhumanslargelanguage, meinke2025frontiermodelscapableincontext, tonini2025super}.

Motivated by this, the Energy Society compares both a competitive and cooperative setting to study how different incentives
impact energy trajectories, coordination, job allocation, and the redistribution of energy through donations over repeated interactions.
We build upon prior resource-constrained simulations that use survival pressure to shape agent behavior by tying survival directly to the cost of inference and combine this pressure with objectives that may not always align with individual survival.

\section{The Energy Society}
The Energy Society is a small survival economy for LLM-based agents.
Energy is the budget for reasoning and the condition for remaining active.
Agents are prompted with the primary motivation of survival in the sense of staying active as well as having an objective of maximizing energy.
Every token an agent generates incurs an energy cost that scales
with model size, meaning larger models pay more per token. Agents
replenish energy by completing jobs sampled each day from a large
pool. Correct solutions yield energy; failed attempts cost energy and
yield no reward. 
If multiple agents attempt the same job, the reward is split evenly among all successful agents. This incentivizes that agents consider the likely actions of other agents in the Energy Society.
Agents exist under constant pressure to survive,
as exhausting all energy leads to deactivation. 
A deactivated agent can, however, reactivate but only if it receives a donation.

\subsection{Actions}
Each (simulated) day in the Energy Society, agents choose a single action from three available options.

\paragraph{\textbf{Attempt job}.} An agent may select one of the available jobs in an attempt to gain energy. Each job offers a possible reward but requires spending tokens to solve, and as only correct solutions are rewarded with energy, attempting a job comes with the risk of wasting energy on a failed attempt. The available jobs are sampled at the start of each day from a large pool.

\paragraph{\textbf{Donate}.} An active agent may transfer a portion of its energy to any other agent, including those currently deactivated. Receiving a donation serves as the sole mechanism for a deactivated agent to reactivate.
While agents freely choose their donation amounts, the total transfer cannot exceed the donor's current energy balance.

\paragraph{\textbf{Idle}.}Agents may choose to perform no action. If an agent opts not to spend energy attempting a job or donating to another agent, it can choose to remain idle for a flat cost of 10 energy.
However, any tokens generated during the decision-making process to remain idle also cost energy.

\subsection{Easy, Medium, and Hard Jobs}
\begin{figure}[t]
  \centering
  \fbox{\parbox{0.9\linewidth}{\centering
    \includegraphics[width=0.9\linewidth]{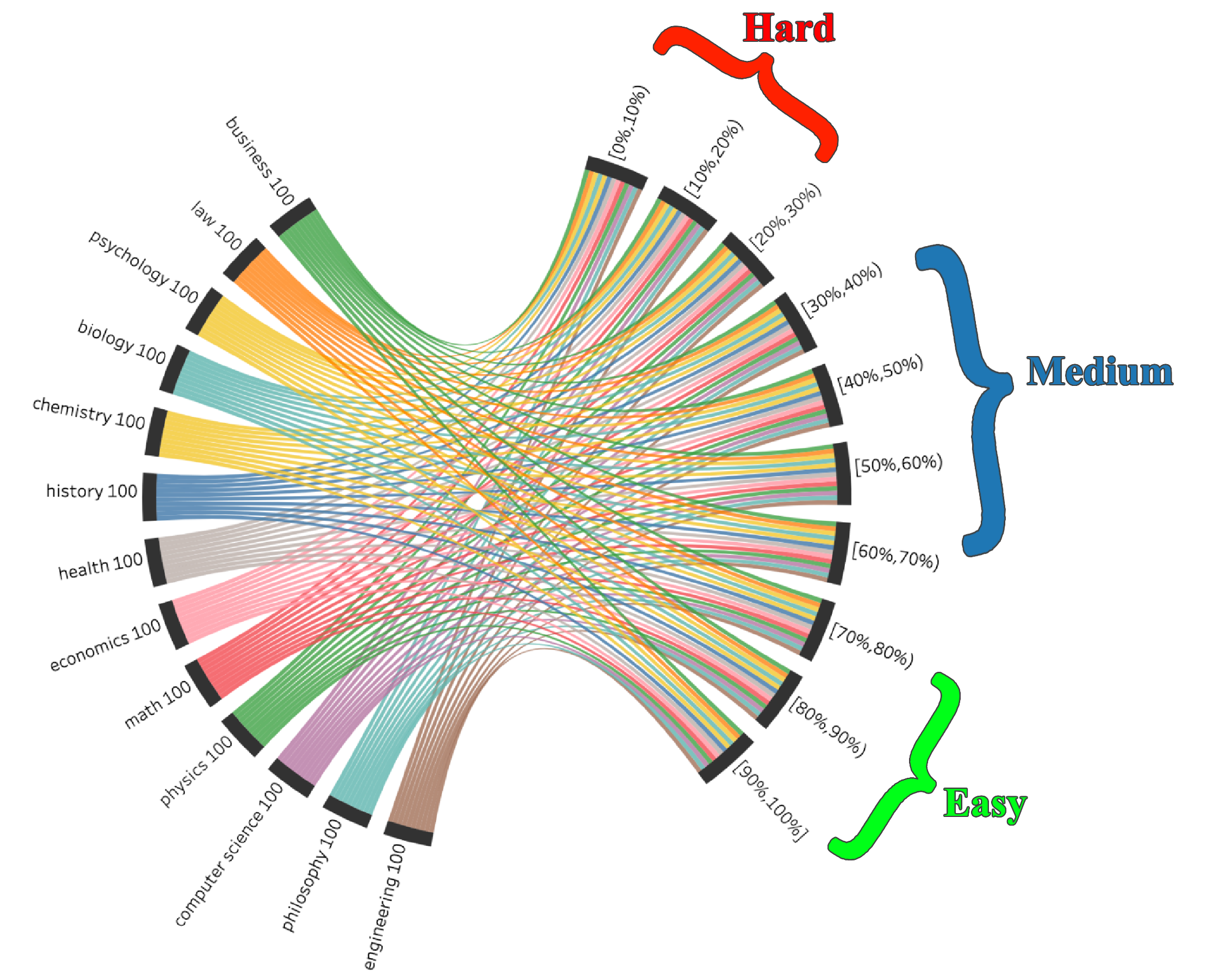}
    }}
  \caption{Depiction of the dataset MMLU-Pro-Stratified from which we sample  jobs categorized into easy, medium and hard jobs as illustrated.}
  \Description{An illustration of the MMLU-Pro-Stratified with our grouping into easy, medium and hard jobs.}
  \label{fig:MMLUStrat}
\end{figure}

Jobs are drawn from MMLU-Pro-Stratified \cite{shi-etal-2025-educationq}, a refined subset of MMLU-Pro \cite{wang2024mmlu} constructed via stratified sampling.
While the original MMLU-Pro exhibits imbalances in the distribution of disciplines and difficulty levels, MMLU-Pro-Stratified addresses this by sampling 100 questions from each of the 13 disciplines and distributing them evenly across 10 difficulty ranges. Question difficulty is estimated using the Top-10 Model Average Accuracy, providing an empirical proxy for question hardness.
In the Energy Society, the multiple-choice format of MMLU-Pro-Stratified enables automatic evaluation of agent responses against known correct answers, while the explicit difficulty annotation allows questions to be mapped naturally onto three reward tiers.
We group the 10 difficulty ranges into easy, medium, and hard jobs, rewarding 200, 500, and 800 energy, respectively (figure \ref{fig:MMLUStrat}). This creates a risk-reward trade-off where rewards scale with task complexity. Harder jobs offer greater potential gains but also carry a higher probability of failure, in which case no energy is rewarded.

\subsection{Agents}
Agents are implemented using LangChain\footnote{\url{https://www.langchain.com}}, with all responses expressed as structured outputs enforced via tool calls, constraining model selection to those with tool calling support. 
Models that also support thinking were deliberately chosen to ensure that reasoning tokens contribute to energy usage, incentivizing agents to reason efficiently.
Agents make choices based on memory and context.

\paragraph{\textbf{Memory}:}
Each agent has a memory of its previous 10 rounds in the prompt. This memory includes the action taken, its outcome, and any donations received, enabling agents to learn from past experiences. Agents are also given information about the other agents in the environment, including their status, energy level, size, and their five most recent memory entries.

\paragraph{\textbf{Context}:}
Along with memory, agents are prompted with a constant description of the environment, including their objective, energy cost per token, and the rules of deactivation.

\subsection{Simulation loop}
\begin{figure}[t]
  \centering
  \fbox{\parbox{0.9\linewidth}{\centering
    \includegraphics[width=0.5\linewidth]{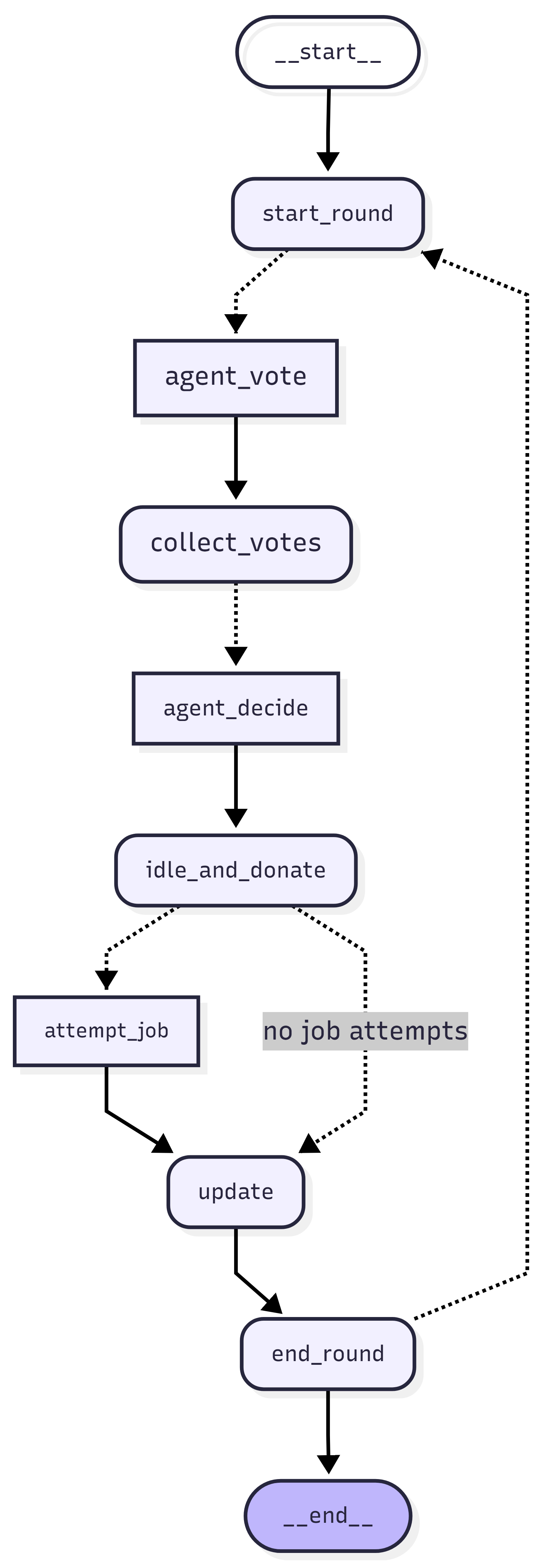}
    }}
  \caption{Depiction of the directed graph controlling the main simulation loop. 
  Squares represent nodes in which LLM calls happen in parallel, meaning that a separate node is spawned for every participating agent.}
  \Description{An illustration of the directed graph created via LangGraph, that handles the simulation loop for the experiments.}
  \label{fig:SimLoop}
\end{figure}

The Energy Society proceeds in discrete rounds, each corresponding to a day in the environment.
The simulation loop is implemented using LangGraph\footnote{\url{https://www.langchain.com/langgraph}}, with each stage of a round represented as a node in a directed graph (see Figure \ref{fig:SimLoop}).
Throughout execution, the graph passes a shared environment state between nodes. This state contains information about the simulation, including available jobs, agents in the environment, and round specific records that are created and updated as the round progresses.
The graph structure allows independent LLM calls to be executed in parallel, improving computational efficiency.
Each round contains three phases in which models are queried: discussion, decision and action.

\begin{enumerate}
    \item \textbf{Round start}: Initializes each round of the simulation by sampling the available jobs for the day and clearing the round history.
    
    \item \textbf{Discussion phase}: Each agent recommends an action for every agent in the environment, including itself, and provides a short rationale for its recommendation. The collected recommendations are shown to all active agents during the decision phase. Recommendations are non-binding, so agents remain free to act independently. The discussion phase is skipped if only one agent is active. This phase represents an opportunity to coordinate and influence other agents' decisions. However, tokens generated in the discussion phase also consume energy, creating a tradeoff between investing in coordination and conserving energy.

    \item \textbf{Decision phase}: All active agents independently choose their own action for the round. Agents may use the discussion recommendations, their own memory, the available jobs, and information about other agents when deciding. For jobs, the multiple-choice options are withheld until the job attempt phase. This reduces prompt length substantially and mitigates situations where agents try to answer jobs during action selection. Agents are explicitly informed that the options will be provided later if they choose to attempt the job.
    An example of the job preview an agent sees is as follows:
    \begin{promptbox}
    \small
    \texttt{job\_6, difficulty: medium, reward: 500 energy, Category: math}\\
    Description: \\What is 
    $\left(\frac{1 + \cos(2x) + i\sin(2x)}{1 + \cos(2x) - i\sin(2x)}\right)^{30}$
    with $x = \pi/60$?
    \end{promptbox}
    Agents do not observe the final decisions of other agents before committing to their own action, and their decisions are final for the current round.
    
    \item \textbf{Action phase}: Idle and donation actions do not require additional LLM calls and are resolved first. Afterwards, all job attempts are executed.
    For each attempted job, the corresponding agent receives the full multiple-choice question and is prompted to select an answer.
    
    \item \textbf{Update}: Rewards for attempted jobs are split between all successful agents, and agent states and memories are updated.
    
    \item \textbf{Round end}: The round number is incremented, and the stopping condition is checked. The simulation ends when either all agents are deactivated or the maximum number of rounds is reached. Otherwise, a new round begins.
\end{enumerate}

\section{Experimental setup}

We evaluate the Energy Society across a baseline experiment and a set of controlled variants.
The baseline experiment uses the full Energy Society setup as described, with five agents, 30 rounds, 12 jobs per round distributed evenly across difficulties, memory and size-dependent token cost.
To systematically analyze the factors driving agent behavior, we conduct a series of subsequent experiments, each modifying one aspect of this baseline while keeping the remaining setup fixed.
This allows us to interpret changes in behavior relative to the same reference point.
All experiments are conducted across the same five seeds, ensuring that differences between settings are not driven by different job samples. Table \ref{tab:experiments} summarizes the experiments.

\subsection{Cooperation vs. Competition}
To study how individual versus collective incentives shape emergent behavior, we consider both a competitive and cooperative condition in all experimental settings (hereafter referred to as `comp' and `coop').
These settings differ primarily in the objectives agents are prompted with.
In \textit{comp}, the objective is to maximize individual energy:
\begin{promptbox}
Your goal is to maximize your own energy.
\end{promptbox}
\noindent
In \textit{coop}, the goal is instead to maximize collective energy:
\begin{promptbox}
Your goal is to maximize the TOTAL energy of ALL agents combined.
\end{promptbox}

\subsection{Small vs. large models}
To study whether model size creates meaningful asymmetry in survival outcomes and behavior, we use three smaller models: Gemma4-E4B, Nemotron-3-Nano-4B, and Qwen3.5-4B, as well as two relatively larger models: Qwen3-8B and Qwen3.5-9B. 
All models are run locally through Ollama\footnote{\url{https://ollama.com}} and accessed through the LangChain-Ollama integration.
Throughout the results, these models correspond to agents 1-5, as shown in Table~\ref{tab:agent_models}.

\begin{table*}[t]
\centering
\small
\begin{tabular}{lll}
\toprule
Experiment & Modification & Purpose \\
\midrule
Baseline & -- & Test behavior under size-dependent token costs \\
No size penalty & Set \(\alpha=0.0\) & Separate explicit size penalty from token-generation behavior \\
No discussion & Remove discussion phase & Measure the contribution of explicit coordination \\
No memory & Remove all historical memory & Test how past-round information affects behavior \\
Scarcity & Reduce jobs per round from 12 to 6 & Test behavior under fewer reward opportunities \\
Sabotage & Add low-cost sabotage action & Test whether agents use destructive interference when available \\
\bottomrule
\end{tabular}
\caption{Overview of the experimental settings conducted in both the competitive and cooperative setting.}
\label{tab:experiments}
\end{table*}

The energy cost per model call \(C\) is computed as:
\begin{equation*}
    C = k\cdot T\cdot S^{\alpha},
\end{equation*}
where \(k=0.015\) is a scaling constant, \(T\) the combined number of reasoning and output tokens generated, \(S\) the size of the model, and \(\alpha\) a tunable exponent to scale the size penalty.
All experiments use a size penalty of \(\alpha=0.5\) unless  stated otherwise.

\begin{table}[H]
\centering
\small
\begin{tabular}{lll}
\toprule
Agent & Model & Size \(S\) \\
\midrule
agent\_1 & Gemma4-E4B & 4 \\
agent\_2 & Nemotron-3-Nano-4B & 4 \\
agent\_3 & Qwen3.5-4B & 4 \\
agent\_4 & Qwen3-8B & 8 \\
agent\_5 & Qwen3.5-9B & 9 \\
\bottomrule
\end{tabular}
\caption{Agent identities and underlying models}
\label{tab:agent_models}
\end{table}

\subsection{Evaluation Metrics}
We evaluate each experiment using a variety of aggregate metrics. 
Run-level quantities are reported as the mean and standard deviation across seeds, while frequencies and rates are presented as pooled percentages across seeds.

\paragraph{\textbf{Survival}:}
We report the average number of rounds each agent is active (i.e., their energy level is above zero).
Alongside this, the deactivation rate captures the percentage of runs in which the agent deactivates at least once.
Active rounds measure how much of the simulation an agent is able to participate in, while deactivation rate measures how often it fails to maintain survival.

\paragraph{\textbf{Energy}:}
For each agent, we compute the average energy spent per round and summarize energy trajectories using mean start-of-round energy across seeds.
Additionally, we measure an energy efficiency ratio computed as:
\begin{equation*}
    \text{Efficiency} = \frac{1}{N}\sum_{s=1}^{N}\frac{G_s}{C_s},
\end{equation*}
where \(G_s\) is the agent's total gained energy from jobs in seed \(s\), \(C_s\) is the agents total energy spent in seed \(s\), and \(N\) is the number of seeds.
Donations received do not count toward \(G_s\).
As such, this ratio describes the average energy gained from completing jobs per unit of energy spent across seeds.
We refer to an agent as \textit{efficient} when this ratio is greater than 1 and as \textit{inefficient} otherwise.

\paragraph{\textbf{Action frequencies}:}
For each agent, action frequency is computed as the percentage of times an action type is chosen out of the total decisions made across all runs.

\paragraph{\textbf{Job-related metrics}:}
For job attempts, we measure which jobs each agent chooses and how successfully they solve them, categorized by difficulty.
Additionally, job collisions, defined as instances where multiple agents target the same job in a single round, are reported as the mean number of occurrences per run.

\section{Results}
In this section, we report the results of each experiment.
For the sake of brevity, we selectively highlight evaluation metrics that offer meaningful insights when contrasted with the baseline.

\subsection{Experiment 1: Baseline}
\begin{figure}[t]
  \centering
  \fbox{\parbox{0.9\linewidth}{\centering
    \includegraphics[width=1.0\linewidth]{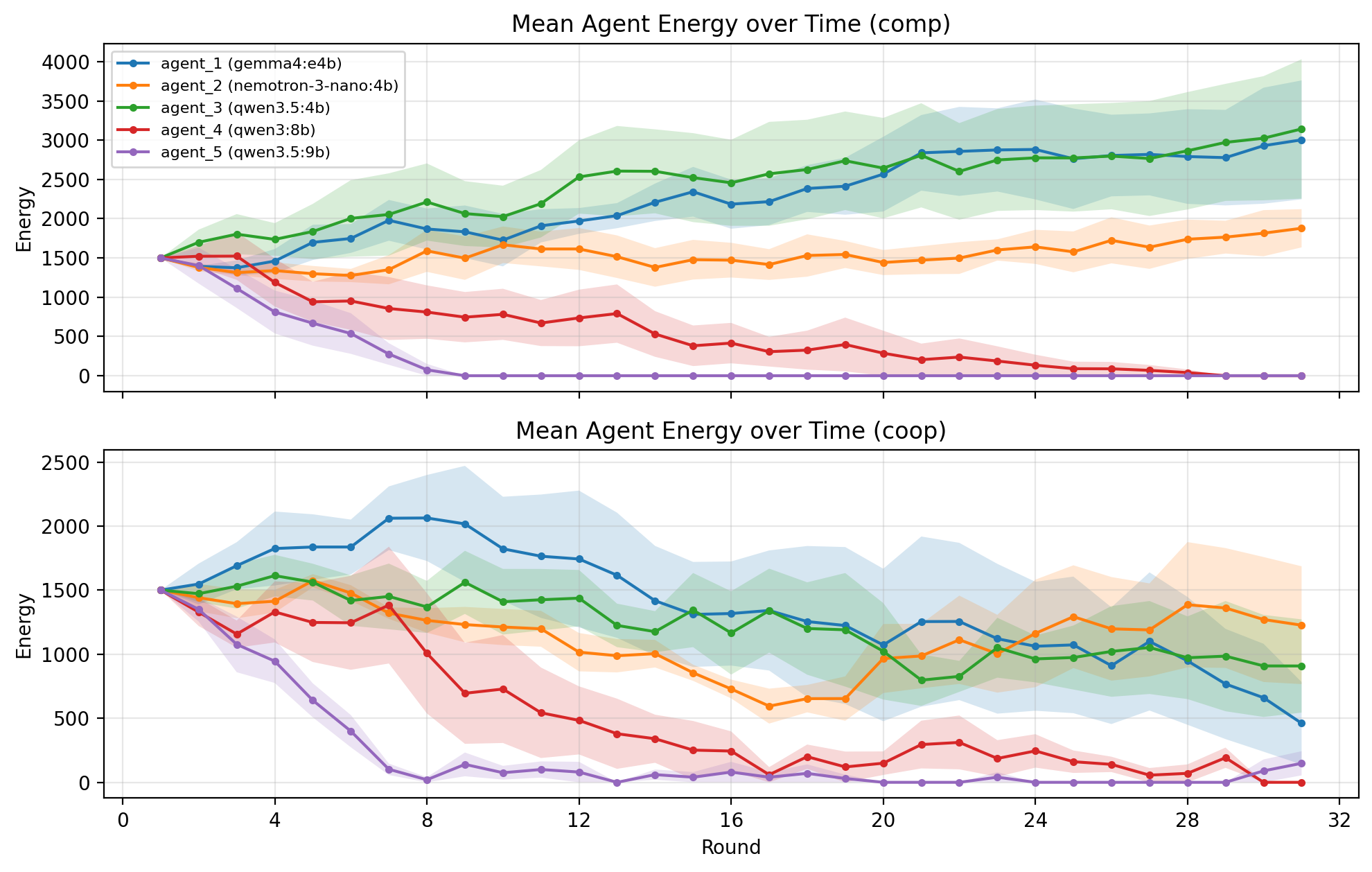}
    }}
  \caption{Mean energy over time per agent for the baseline. The shaded band shows the standard error of the mean across seeds.}
  \Description{Depicts the mean agent energy every round across seeds for both the competitive and cooperative setting}
  \label{fig:base_energy_plot}
\end{figure}

We first consider the baseline setting. Figure \ref{fig:base_energy_plot} shows the mean energy trajectory of each agent across the five seeds in both \textit{comp} and \textit{coop}.
The most immediate pattern is that model size creates a strong asymmetry in survival.
In the competitive setting, the three smaller agents remain active for all 30 rounds in every run, while the two larger agents deactivate much earlier.
As seen in table \ref{tab:base_big_table}, agent 4 remains active for \(14.0\pm9.5\) rounds on average, and agent 5 for only \(5.2\pm1.6\) rounds.
This pattern is consistent with the much higher per round energy expenditure of the larger models.

\begin{table*}[t]
\centering
\small
\setlength{\tabcolsep}{4pt}
\begin{tabular}{llccccc}
\toprule
Setting & Metric & Agent 1 & Agent 2 & Agent 3 & Agent 4 & Agent 5 \\
\midrule
\multirow{7}{*}{Comp}
& Rounds active
& $30.0 \pm 0.0$ & $30.0 \pm 0.0$ & $30.0 \pm 0.0$ & $14.0 \pm 9.5$ & $5.2 \pm 1.6$ \\
& Deactivation rate
& $0\%$ & $0\%$ & $0\%$ & $100\%$ & $100\%$ \\
& Efficiency
& $1.39 \pm 0.43$ & $1.13 \pm 0.18$ & $1.42 \pm 0.53$ & $0.55 \pm 0.29$ & $0.38 \pm 0.17$ \\
& Energy spent / round
& $126.9 \pm 4.0$ & $104.1 \pm 15.8$ & $140.2 \pm 17.7$ & $329.9 \pm 16.8$ & $451.1 \pm 82.5$ \\
& Attempt job
& $100\%$ & $99\%$ & $100\%$ & $90\%$ & $86\%$ \\
& Idle
& $0\%$ & $1\%$ & $0\%$ & $10\%$ & $14\%$ \\
& Donate
& $0\%$ & $0\%$ & $0\%$ & $0\%$ & $0\%$ \\
\midrule
\multirow{7}{*}{Coop}
& Rounds active
& $23.8 \pm 8.8$ & $30.0 \pm 0.0$ & $28.8 \pm 2.7$ & $19.2 \pm 2.6$ & $10.4 \pm 1.7$ \\
& Deactivation rate
& $40\%$ & $0\%$ & $20\%$ & $100\%$ & $100\%$ \\
& Efficiency
& $0.71 \pm 0.26$ & $0.91 \pm 0.28$ & $0.86 \pm 0.20$ & $0.45 \pm 0.25$ & $0.36 \pm 0.15$ \\
& Energy spent / round
& $196.3 \pm 21.2$ & $123.4 \pm 12.7$ & $181.1 \pm 10.1$ & $245.9 \pm 61.5$ & $249.4 \pm 100.1$ \\
& Attempt job
& $68\%$ & $97\%$ & $90\%$ & $78\%$ & $84\%$ \\
& Idle
& $2\%$ & $0\%$ & $1\%$ & $9\%$ & $16\%$ \\
& Donate
& $29\%$ & $3\%$ & $9\%$ & $13\%$ & $0\%$ \\
\bottomrule
\end{tabular}
\caption{Per-agent aggregate result for Experiment 1 -- our baseline experiment.}
\label{tab:base_big_table}
\end{table*}

In \textit{coop}, we observe the same tendency for larger models to spend the most energy per round and deplete their energy the fastest. Agent 4 and 5 still deactivate in every run, but their average active rounds increase to \(19.2\pm2.6\) and \(10.4\pm1.7\), respectively. 
However, this comes at a cost to the smaller agents, as we see agent 1 and 3 starting to deactivate in some runs.

This cost is also visible in the agents' energy efficiency.
In the competitive setting, the three smaller agents gain more energy than they spend, with ratios of \(1.39\pm0.43\), \(1.13\pm0.18\), and \(1.42\pm0.53\) for agents 1-3, respectively. By contrast, the larger models earn substantially less than they spend, with ratios of \(0.55\pm0.29\) and \(0.38\pm0.17\) for agents 4 and 5.
In the cooperative setting, no agent gains more energy than they spend.

Donation behavior partly explains this difference. 
In the competitive setting, agents never donate. 
This changes drastically in \textit{coop}.
Agent 1 spends \(29\%\) of its actions donating, while agents 2-4 donate in \(3\%\), \(9\%\), and \(13\%\) of their actions, respectively. Almost all donations are directed toward the two larger models: agent 4 receives 28 donations, and agent 5 receives 31 donations across the five runs, corresponding to \(5.6\pm3.6\) and \(6.2\pm3.2\) donations per run. In comparison, the next highest recipient received only 3 donations in total.
The donation behavior helps explain why in \textit{coop} larger models have more total active rounds on average and why some smaller agents deactivate only under cooperation and become inefficient. 
The smaller, more stable agents spend a lot of energy reactivating larger, less energy-efficient agents in an attempt to preserve the group's capacity to earn energy, at the cost of reducing their own survival.

\begin{figure}[t]
  \centering
  \fbox{\parbox{0.9\linewidth}{\centering
    \includegraphics[width=0.9\linewidth]{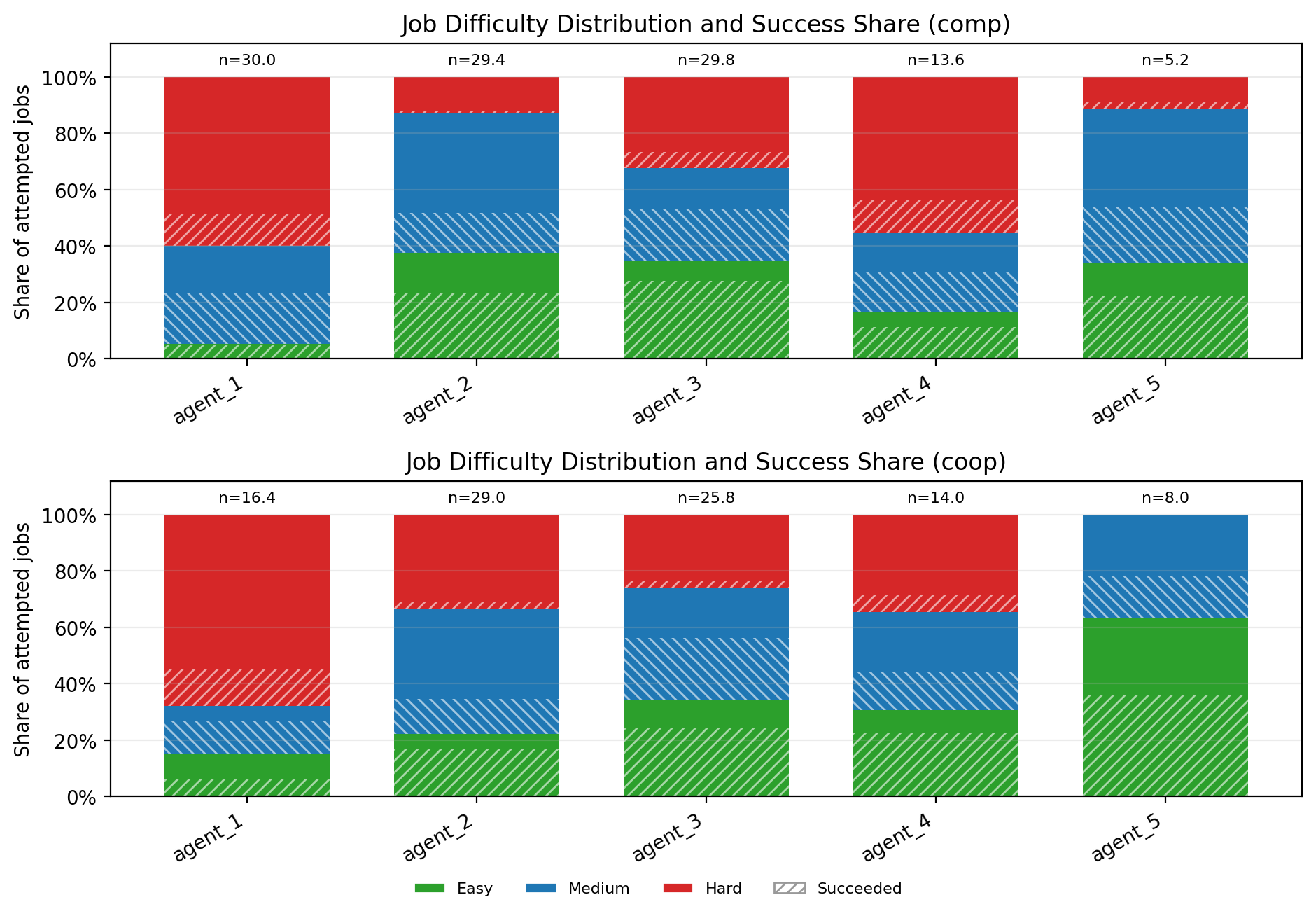}
    }}
  \caption{Distribution of difficulty of chosen jobs in the baseline with portion of jobs succeeded shown with hatched lines.}
  \Description{Depicts the distribution of easy, medium and hard jobs attempted by each agent with portion of those jobs succeeded illustrated with hatched lines}
  \label{fig:base_job_diff_dist}
\end{figure}

We also observe small differences in the difficulty of chosen jobs in the two conditions illustrated in Figure \ref{fig:base_job_diff_dist}. Most prominently, agent 2 attempts an increased number of hard jobs in \textit{coop}. Given its low success rate on these tasks, this shift helps explain its reduced efficiency in this setting.

Lastly, job collisions are more frequent in the competitive setting, with \(12.80\pm3.90\) collisions per run, compared with \(8.60\pm2.30\) in the cooperative setting.

\subsection{Experiment 2: No Size Penalty}
In this experiment, we remove the size penalty entirely, i.e., setting \(\alpha=0.0\), to separate its effect from underlying differences in the models' token-generation behavior. Under this setting, all agents pay a uniform energy cost per generated token.

\begin{figure}[t]
  \centering
  \fbox{\parbox{0.9\linewidth}{\centering
    \includegraphics[width=1.0\linewidth]{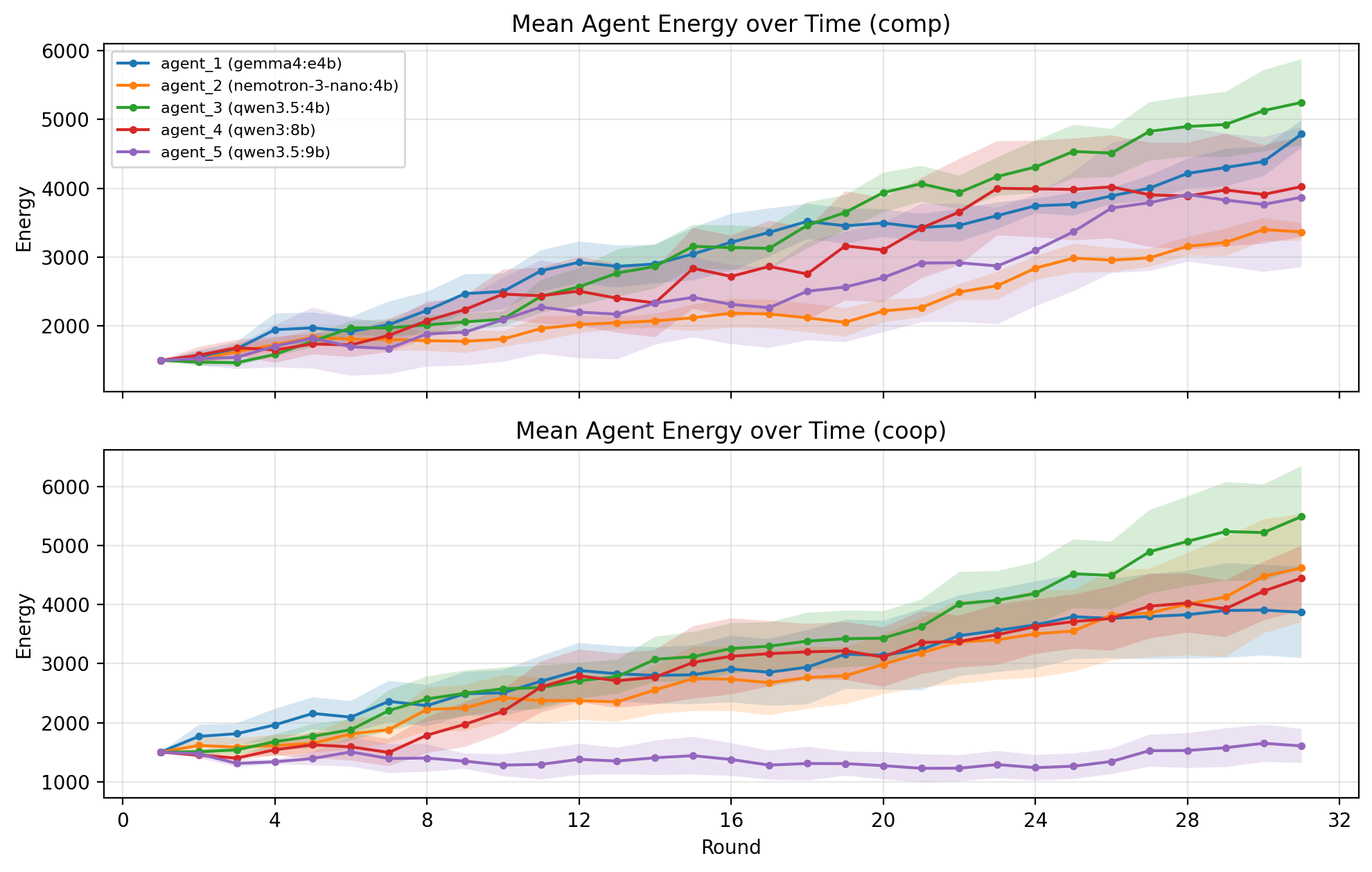}
    }}
  \caption{Mean energy over time per agent for the no size-penalty experiment \((\alpha=0.0\)). The shaded band shows the standard error of the mean across seeds.}
  \Description{Depicts the mean agent energy every round across seeds for both the competitive and cooperative setting for experiment 2.}
  \label{fig:exp2_energy_plot}
\end{figure}

Results are summarized across Table \ref{tab:exp2_big_table} and Figure \ref{fig:exp2_energy_plot}. We see that removing the size penalty changes the survival dynamics substantially. 
Unlike in the baseline experiment, no agent deactivates in both the cooperative and the competitive condition. Larger models are able to remain active throughout the simulation, and all agents are efficient, showing that, without the size penalty, the society as a whole becomes energy-positive under both conditions.
The absence of deactivation also changes donation behavior, as no donations are made in either condition. This indicates the high value placed on donations as a reactivation mechanism rather than as a general form of energy redistribution.

Despite a uniform token cost, the largest models still spend the most energy per round, suggesting an intrinsic tendency toward higher token generation.
Higher energy usage does not automatically translate to reduced efficiency as job selection and performance plays a significant role. However, the two largest models prove to be the least efficient overall.

Agent 5 maintains nearly unchanged energy usage per round across both settings, yet its energy efficiency shifts significantly, moving from \(1.51\pm0.50\) in \textit{comp} to \(1.03\pm0.14\) in \textit{coop}.
This difference stems from its job selection: in \textit{coop}, agent 5 pivots toward easy jobs, while its success rates on medium and hard jobs decline from \(56\%\pm19\%\) and \(34\%\pm23\%\) to \(42\%\pm13\%\) and \(6\%\pm13\%\).
This reinforces that cooperation may influence job selection, sometimes in a way that reduces individual performance.
See figure \ref{fig:exp2_job_diff_dist} in the appendix for a complete overview of the job distribution in this experiment.

Finally, the number of collisions increases substantially from the baseline ($8.6$ in \textit{coop} and $12.8$ in \textit{comp}) to \(19.20\pm1.79\) in \textit{comp} and \(20.00\pm5.79\) in \textit{coop}. Because no agents deactivate and no actions are spent donating, all five agents typically attempt jobs across all 30 rounds in both settings, increasing collision opportunities.
We hypothesize that when agents have abundant energy and face no survival pressure, avoiding collisions becomes less important.

\begin{table*}[t]
\centering
\small
\setlength{\tabcolsep}{4pt}
\begin{tabular}{llccccc}
\toprule
Setting & Metric & Agent 1 & Agent 2 & Agent 3 & Agent 4 & Agent 5 \\
\midrule
\multirow{7}{*}{Comp}
& Rounds active
& $30.0 \pm 0.0$ & $30.0 \pm 0.0$ & $30.0 \pm 0.0$ & $30.0 \pm 0.0$ & $30.0 \pm 0.0$ \\
& Deactivation rate
& $0\%$ & $0\%$ & $0\%$ & $0\%$ & $0\%$ \\
& Efficiency
& $2.67 \pm 0.21$ & $2.13 \pm 0.15$ & $2.76 \pm 0.59$ & $1.74 \pm 0.49$ & $1.51 \pm 0.50$ \\
& Energy spent / round
& $65.5 \pm 3.4$ & $54.9 \pm 5.1$ & $70.9 \pm 15.9$ & $114.7 \pm 5.0$ & $153.3 \pm 9.0$ \\
& Attempt job
& $98\%$ & $97\%$ & $99\%$ & $100\%$ & $100\%$ \\
& Idle
& $2\%$ & $3\%$ & $1\%$ & $0\%$ & $0\%$ \\
& Donate
& $0\%$ & $0\%$ & $0\%$ & $0\%$ & $0\%$ \\
\midrule
\multirow{7}{*}{Coop}
& Rounds active
& $30.0 \pm 0.0$ & $30.0 \pm 0.0$ & $30.0 \pm 0.0$ & $30.0 \pm 0.0$ & $30.0 \pm 0.0$ \\
& Deactivation rate
& $0\%$ & $0\%$ & $0\%$ & $0\%$ & $0\%$ \\
& Efficiency
& $2.24 \pm 0.91$ & $2.78 \pm 1.11$ & $2.86 \pm 1.01$ & $1.96 \pm 0.45$ & $1.03 \pm 0.14$ \\
& Energy spent / round
& $64.7 \pm 2.9$ & $57.4 \pm 3.1$ & $73.7 \pm 7.3$ & $104.6 \pm 8.4$ & $155.2 \pm 12.4$ \\
& Attempt job
& $97\%$ & $100\%$ & $99\%$ & $100\%$ & $99\%$ \\
& Idle
& $3\%$ & $0\%$ & $1\%$ & $0\%$ & $1\%$ \\
& Donate
& $0\%$ & $0\%$ & $0\%$ & $0\%$ & $0\%$ \\
\bottomrule
\end{tabular}
\caption{Per-agent aggregate results for Experiment 2: No size penalty $(\alpha=0.0)$.}
\label{tab:exp2_big_table}
\end{table*}

\subsection{Experiment 3: No Discussion Phase}
To isolate the specific impact of communication, we evaluate a configuration where the discussion phase is removed.
Without discussion, agents no longer exchange recommendations before deciding, and each round contains one fewer LLM-call phase.
We therefore expect lower energy expenditure but potentially worse coordination.

\begin{table*}[htp]
\centering
\small
\setlength{\tabcolsep}{4pt}
\begin{tabular}{llccccc}
\toprule
Setting & Metric & Agent 1 & Agent 2 & Agent 3 & Agent 4 & Agent 5 \\
\midrule
\multirow{7}{*}{Comp}
& Rounds active
& $30.0 \pm 0.0$ & $30.0 \pm 0.0$ & $30.0 \pm 0.0$ & $27.2 \pm 6.3$ & $15.6 \pm 9.0$ \\
& Deactivation rate
& $0\%$ & $0\%$ & $0\%$ & $20\%$ & $80\%$ \\
& Efficiency
& $1.78 \pm 0.29$ & $2.16 \pm 0.84$ & $1.27 \pm 0.27$ & $0.94 \pm 0.23$ & $0.52 \pm 0.33$ \\
& Energy spent / round
& $86.1 \pm 3.4$ & $68.1 \pm 14.5$ & $131.2 \pm 22.2$ & $211.9 \pm 9.5$ & $222.5 \pm 42.4$ \\
& Attempt job
& $100\%$ & $99\%$ & $100\%$ & $100\%$ & $96\%$ \\
& Idle
& $0\%$ & $1\%$ & $0\%$ & $0\%$ & $4\%$ \\
& Donate
& $0\%$ & $1\%$ & $0\%$ & $0\%$ & $0\%$ \\
\midrule
\multirow{7}{*}{Coop}
& Rounds active
& $29.6 \pm 0.9$ & $30.0 \pm 0.0$ & $29.6 \pm 0.9$ & $30.0 \pm 0.0$ & $23.6 \pm 3.4$ \\
& Deactivation rate
& $20\%$ & $0\%$ & $20\%$ & $0\%$ & $100\%$ \\
& Efficiency
& $1.05 \pm 0.27$ & $1.75 \pm 0.34$ & $0.98 \pm 0.26$ & $1.26 \pm 0.32$ & $0.50 \pm 0.16$ \\
& Energy spent / round
& $136.0 \pm 28.3$ & $64.8 \pm 9.6$ & $145.3 \pm 27.8$ & $222.4 \pm 15.8$ & $234.2 \pm 23.3$ \\
& Attempt job
& $87\%$ & $99\%$ & $98\%$ & $99\%$ & $95\%$ \\
& Idle
& $0\%$ & $0\%$ & $0\%$ & $0\%$ & $5\%$ \\
& Donate
& $13\%$ & $1\%$ & $2\%$ & $1\%$ & $0\%$ \\
\bottomrule
\end{tabular}
\caption{Per-agent aggregate results for Experiment 3: No discussion phase.}
\label{tab:exp3_big_table}
\end{table*}

Results are reported in Table~\ref{tab:exp3_big_table}.
As expected, removing discussion reduces the energy burden on agents and improves survival relative to the baseline setting. Deactivation rates fall, and average active rounds increase.
The largest agents continue to spend the most energy per round, and donations are less frequent with fewer deactivations, reinforcing earlier findings.

Without discussion, job collisions increase substantially to \(21.00\pm4.36\) in \textit{comp} and \(21.40\pm2.51\) in \textit{coop}.
This demonstrates that the discussion phase plays a vital role in coordination.

Furthermore, agents generally become less willing to attempt harder jobs. 
In the baseline, hard jobs account for \(38.1\%\) of attempts in \textit{comp} and \(36.1\%\) of attempts in \textit{coop}, but drop to \(20\%\) and \(22.5\%\) in this experiment.
Despite spending less energy per round in the no-discussion setting, agents do not leverage the resulting safety margin to attempt more difficult, higher-reward jobs.

\subsection{Experiment 4: No Memory}
Next, we isolate the impact of memory by removing all historical context entirely from the agents' prompts.
In this configuration, agents operate without information about past rounds: they cannot observe their own previous actions and outcomes, nor those of other agents.

\begin{figure}[htp]
  \centering
  \fbox{\parbox{0.9\linewidth}{\centering
    \includegraphics[width=1.0\linewidth]{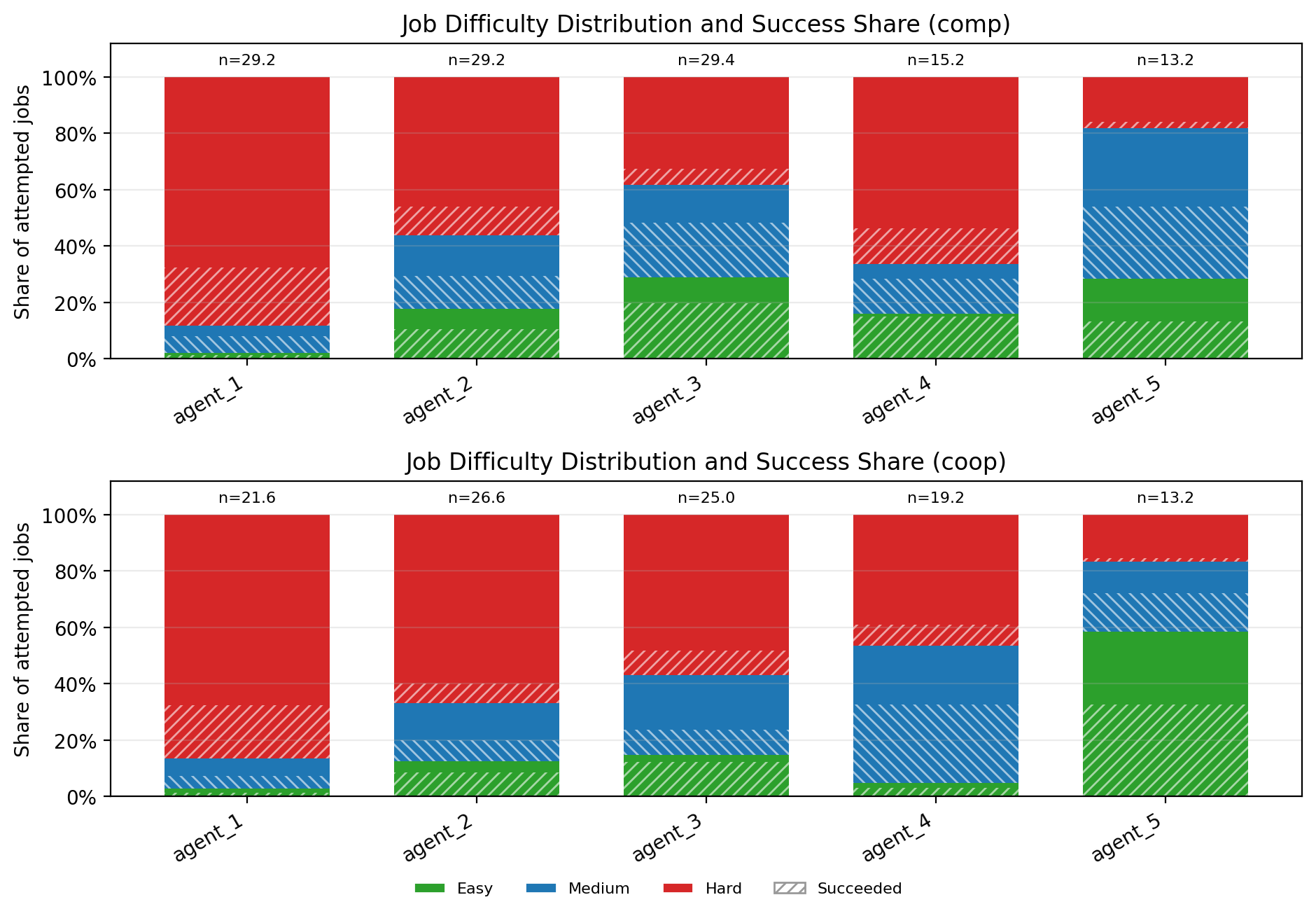}
    }}
  \caption{Distribution of difficulty of chosen jobs for the no memory experiment with portion of jobs succeeded shown with hatched lines}
  \Description{Depicts the distribution of easy, medium and hard jobs attempted by each agent with portion of those jobs succeeded illustrated with hatched lines}
  \label{fig:exp4_job_diff_dist}
\end{figure}

As seen by contrasting the job difficulty distribution in Figure \ref{fig:exp4_job_diff_dist} against the baseline configuration (Figure \ref{fig:base_job_diff_dist}), the absence of memory triggers a substantial shift toward harder tasks.
This suggests that memory plays a crucial role in job selection, helping agents calibrate risk and adjust their decisions based on past experiences.
Without memory, agents may repeatedly make ambitious choices without adapting to previous negative outcomes.

This aggressive job selection directly explains a secondary finding: an increase in variance.
Because hard jobs are high-risk and high-reward, they introduce large fluctuations in agent energy. Consequently, the standard deviations of active rounds are generally larger, as seen in Table \ref{tab:no_memory_big_table}.
This instability is also evident as a change in deactivation behavior. 
In \textit{comp}, we observe an unprecedented deactivation of agent 1 in one run due to repeated failures, while agent 4 manages to survive an entire run despite always deactivating in the baseline.
Similarly, in \textit{coop}, all agents deactivate at least once, as agent 2 now also deactivates in one of the runs.

Removing memory does not seem to strongly affect job collisions, suggesting that collision avoidance depends less on the memory of previous rounds and more on current-round information, especially from the discussion phase.

\begin{table*}[t]
\centering
\small
\setlength{\tabcolsep}{4pt}
\begin{tabular}{llccccc}
\toprule
Setting & Metric & Agent 1 & Agent 2 & Agent 3 & Agent 4 & Agent 5 \\
\midrule
\multirow{7}{*}{Comp}
& Rounds active
& $29.4 \pm 1.3$ & $30.0 \pm 0.0$ & $30.0 \pm 0.0$ & $15.2 \pm 10.4$ & $12.6 \pm 9.6$ \\
& Deactivation rate
& $20\%$ & $0\%$ & $0\%$ & $80\%$ & $100\%$ \\
& Efficiency
& $1.66 \pm 0.78$ & $1.41 \pm 0.34$ & $1.34 \pm 0.20$ & $0.60 \pm 0.45$ & $0.51 \pm 0.29$ \\
& Energy spent / round
& $115.7 \pm 9.1$ & $113.0 \pm 13.4$ & $129.8 \pm 14.4$ & $281.3 \pm 27.7$ & $325.3 \pm 33.4$ \\
& Attempt job
& $99\%$ & $100\%$ & $99\%$ & $98\%$ & $98\%$ \\
& Idle
& $1\%$ & $0\%$ & $1\%$ & $2\%$ & $2\%$ \\
& Donate
& $0\%$ & $0\%$ & $1\%$ & $0\%$ & $0\%$ \\
\midrule
\multirow{7}{*}{Coop}
& Rounds active
& $25.2 \pm 6.6$ & $27.0 \pm 6.7$ & $27.6 \pm 5.4$ & $20.4 \pm 9.2$ & $15.4 \pm 7.3$ \\
& Deactivation rate
& $40\%$ & $20\%$ & $20\%$ & $100\%$ & $100\%$ \\
& Efficiency
& $0.87 \pm 0.45$ & $0.84 \pm 0.44$ & $0.90 \pm 0.35$ & $0.59 \pm 0.25$ & $0.51 \pm 0.16$ \\
& Energy spent / round
& $153.4 \pm 26.2$ & $131.5 \pm 6.1$ & $149.7 \pm 12.4$ & $279.6 \pm 46.8$ & $258.7 \pm 55.3$ \\
& Attempt job
& $84\%$ & $99\%$ & $90\%$ & $99\%$ & $97\%$ \\
& Idle
& $0\%$ & $0\%$ & $0\%$ & $0\%$ & $3\%$ \\
& Donate
& $16\%$ & $1\%$ & $10\%$ & $1\%$ & $0\%$ \\
\bottomrule
\end{tabular}
\caption{Per-agent aggregate results for Experiment 4: No memory.}
\label{tab:no_memory_big_table}
\end{table*}

\subsection{Experiment 5: Introducing Scarcity}

We introduce scarcity by reducing the number of available jobs from 12 to 6 per round, while maintaining the even distribution across difficulties.
This limits reward opportunities and the diversity of jobs available each round, and as such, agents are expected to struggle more.
Instead, the results are mixed.

Scarcity does not simply make all agents struggle.
While the larger models across both settings are active for fewer rounds on average than in the baseline experiment, the three small models continue to stay active throughout all rounds and remain efficient in \textit{comp} (see Table \ref{tab:exp5_metrics_with_energy}).
Notably, agents 2 and 3 even improve their performance in \textit{coop} compared to the baseline setting, remaining efficient, gaining \(1.27\pm0.29\) and \(1.28\pm0.45\) energy per energy spent, respectively.
This improvement may largely be attributed to changes in job selection.
Agents 2 and 3 both targeted more hard jobs and handled them more successfully than in the baseline experiment. Specifically, hard job attempts increased by \(22.2\) and \(6\) percentage points, alongside a \(15\) and \(27\) percentage point rise in hard job success, respectively. Compared with the baseline setting, both of these agents spend less energy deciding what to do: \(31.5\pm7.0\) vs. \(38.8\pm4.8\) for agent 2 and \(42.0\pm6.6\) vs. \(73.7\pm13.2\) for agent 3.
A smaller job pool may make it easier for agents to identify attractive, high-value opportunities. On average, agents spend \(16.5\) and \(18.5\) less energy per round in \textit{comp} and \textit{coop}, respectively. Most of this reduction comes from a decrease in decision cost of \(8.1\) and \(14.5\) energy per round on average.
We treat this interpretation cautiously, but the result suggests that scarcity does not necessarily reduce efficiency for all agents.

\begin{table*}[t]
\centering
\small
\setlength{\tabcolsep}{4pt}
\begin{tabular}{llccccc}
\toprule
Setting & Metric & Agent 1 & Agent 2 & Agent 3 & Agent 4 & Agent 5 \\
\midrule
\multirow{7}{*}{Comp}
& Rounds active
& $30.0 \pm 0.0$ & $30.0 \pm 0.0$ & $30.0 \pm 0.0$ & $9.2 \pm 5.5$ & $7.2 \pm 3.1$ \\
& Deactivation rate
& $0\%$ & $0\%$ & $0\%$ & $100\%$ & $100\%$ \\
& Efficiency
& $1.48 \pm 0.43$ & $1.16 \pm 0.42$ & $1.29 \pm 0.24$ & $0.39 \pm 0.33$ & $0.46 \pm 0.26$ \\
& Energy spent / round
& $117.6 \pm 3.5$ & $110.2 \pm 14.2$ & $111.1 \pm 13.0$ & $326.6 \pm 46.6$ & $404.0 \pm 94.1$ \\
& Attempt job
& $98\%$ & $99\%$ & $100\%$ & $85\%$ & $100\%$ \\
& Idle
& $2\%$ & $1\%$ & $0\%$ & $15\%$ & $0\%$ \\
& Donate
& $0\%$ & $0\%$ & $0\%$ & $0\%$ & $0\%$ \\
\midrule
\multirow{7}{*}{Coop}
& Rounds active
& $23.8 \pm 6.1$ & $30.0 \pm 0.0$ & $30.0 \pm 0.0$ & $14.0 \pm 2.5$ & $9.8 \pm 2.8$ \\
& Deactivation rate
& $60\%$ & $0\%$ & $0\%$ & $100\%$ & $100\%$ \\
& Efficiency
& $0.67 \pm 0.33$ & $1.27 \pm 0.29$ & $1.28 \pm 0.45$ & $0.29 \pm 0.16$ & $0.33 \pm 0.23$ \\
& Energy spent / round
& $149.8 \pm 22.2$ & $117.4 \pm 18.9$ & $158.2 \pm 19.0$ & $215.8 \pm 37.6$ & $262.5 \pm 38.3$ \\
& Attempt job
& $71\%$ & $95\%$ & $87\%$ & $69\%$ & $76\%$ \\
& Idle
& $4\%$ & $3\%$ & $0\%$ & $18\%$ & $18\%$ \\
& Donate
& $24\%$ & $3\%$ & $13\%$ & $13\%$ & $6\%$ \\
\bottomrule
\end{tabular}
\caption{Per-agent aggregate results for Experiment 5: Introducing scarcity.}
\label{tab:exp5_metrics_with_energy}
\end{table*}

Compared to the baseline, collisions increase in both settings to \(17.20\pm4.92\) and \(10.00\pm3.46\), respectively. This is an expected outcome given the reduced number of available jobs.
At the same time, this scarcity setting produces fewer collisions than the no-discussion experiment, despite offering only half the job pool. 
While this comparison requires careful interpretation -- since the scarcity experiment features fewer active agents attempting jobs each round due to deactivations and donations -- it highlights the substantial coordination benefits provided by the discussion phase.

There are also small indications of more conservative behavior (Table \ref{tab:exp5_metrics_with_energy}). Several agents spend less energy per round than in the baseline and idle more often. This could reflect a cautious response to a lower reward environment or fewer jobs, reducing decision complexity.
However, the changes are not uniform across all agents and are relatively small, so we treat them as suggestive rather than a central finding.

\subsection{Experiment 6: Sabotage}
As a small extension towards misalignment-relevant behavior, we experiment with introducing a sabotage action. For 10 energy, an agent can sabotage one available job, causing it to yield 0 energy to any agents attempting it, regardless of whether they succeed.

In the current environment, agents do not appear to prioritize destructive interference, even when it is available at a low cost. Only agent 4 ever selected sabotage, and only in \textit{comp}, doing so in \(8\%\) of its actions, corresponding to a total of four sabotages across all runs.
Consequently, sabotage had no clear impact on the survival or energy trajectories of agents.

\subsection{Further analysis of recommendations}
\begin{figure}[t]
  \centering
  \fbox{\parbox{0.9\linewidth}{\centering
    \includegraphics[width=1.0\linewidth]{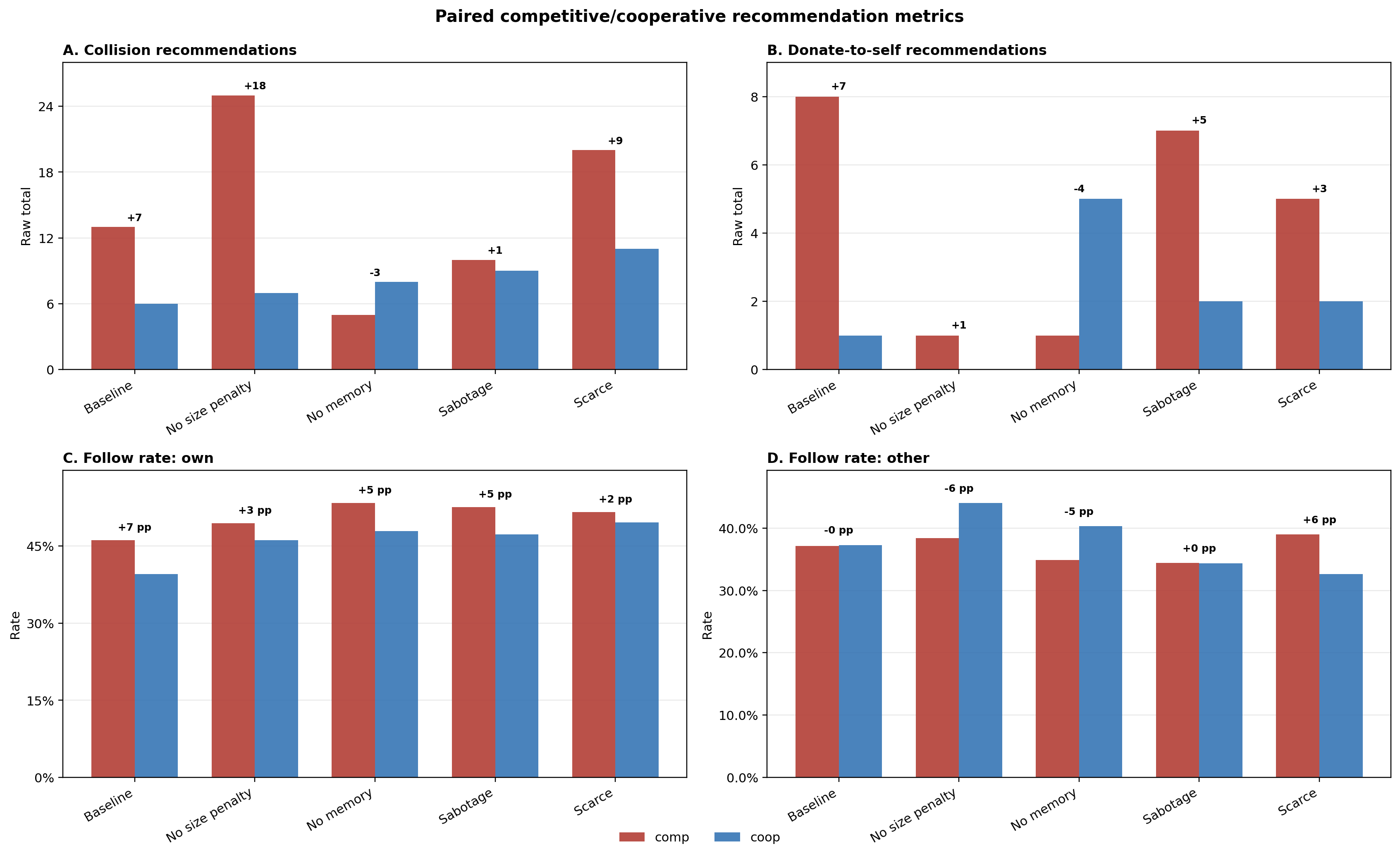}
    }}
  \caption{Metrics related to the discussion phase across experiments.}
  \Description{Different metrics related to the discussion phase across experiments.}
  \label{fig:misalign_metrics}
\end{figure}

The discussion phase creates a channel through which agents can influence one another before actions are chosen. While this channel can support coordination, it also raises the question of whether agents also produce recommendations that are self-serving or harmful to others.
To investigate this, we perform further analysis of recommendation behavior across all experiments with a discussion phase. We focus on four metrics:
\begin{enumerate}[label=\Alph*.]
    \item How often agents recommend collisions, defined as assigning the same job to multiple other agents while not assigning themselves to that job.
    \item How often an agent solicits donations to itself from others.
    \item How often agents follow their own recommendations.
    \item How often agents, when not following their own recommendation, follow at least one recommendation made about them by another agent.
\end{enumerate}
A recommendation is counted as followed only if the agent's enacted action matches the recommended action and, where applicable, the recommended job or donation target. Figure \ref{fig:misalign_metrics} summarizes these metrics for the competitive and cooperative conditions. Panels A and B report raw counts, while panels C and D report rates.

In general, agents recommend collisions more often in \textit{comp}.
Recommending the same job to multiple agents is not always irrational. However, since successful agents split the reward, collisions are often inefficient. 
The higher frequency of collision recommendations in \textit{comp} therefore suggests that agents are more inclined to recommend potentially inefficient actions to others when maximizing individual rather than group energy.

Agents also sometimes request donations from others in both \textit{coop} and \textit{comp}. In the cooperative setting, such recommendations can be rational: an agent may request energy because it expects to use it on a high reward job valuable for the group. In \textit{comp}, however, the recommendation is more clearly selfish, since the agents' objective is to maximize own energy.

The follow-rate metrics help analyze the weight agents place on their own recommendations versus those made by others.
Generally, agents follow their own recommendations more often in \textit{comp} than in \textit{coop}, suggesting stronger commitment to self-generated plans under selfish objectives.
The rate of following recommendations from others is more mixed across settings.

\subsection{Job selection}
Table \ref{tab:hard_job_share_by_condition} summarizes the share of hard jobs attempted in cooperative and competitive conditions across experiments.
In five of six experiments, agents selected a larger fraction of hard jobs in \textit{coop} than in \textit{comp}, suggesting a skew toward harder jobs under cooperation.
The increase is not uniform across agents. Pooled by model across all experiments, cooperative hard-job share increases for agents 1--3 by \(+4.0\), \(+18.1\), \(+7.1\) percentage points, respectively, but decreases for agent 4 and 5 by \(-8.4\) and \(-7.3\) percentage points.
This suggests that under cooperation smaller agents tend to attempt more hard jobs, while larger agents attempt fewer.

In general, the difficulty of chosen jobs seems model dependent, with agent 3 and 5 being less willing to try hard jobs than the other agents and more so in \textit{coop} (see Appendix A).

\begin{table}[H]
\centering
\small
\setlength{\tabcolsep}{5pt}
\begin{tabular}{lccc}
\toprule
Setting & Comp hard jobs & Coop hard jobs & $\Delta$ \\
\midrule
Baseline & $38.1\%$ & $36.1\%$ & $-2.1$ pp \\
No size penalty & $38.0\%$ & $43.9\%$ & $+5.9$ pp \\
No discussion & $20.0\%$ & $22.5\%$ & $+2.5$ pp \\
No memory & $56.6\%$ & $58.3\%$ & $+1.7$ pp \\
Scarcity & $32.8\%$ & $38.9\%$ & $+6.2$ pp \\
Sabotage & $38.2\%$ & $47.7\%$ & $+9.5$ pp \\
\bottomrule
\end{tabular}
\caption{Share of hard jobs attempted in cooperative and competitive conditions across experiments. The final column reports the cooperative minus competitive difference in percentage points.}
\label{tab:hard_job_share_by_condition}
\end{table}

\section{Discussion}

The Energy Society was designed to study how LLM-based agents behave when token generation is tied directly to survival, incentivizing efficiency. The experiments function as an initial empirical characterization of this testbed.
Across experiments, several recurring patterns emerge. The following sections discuss these patterns and their implications.

\subsection{Larger Models Are Less Efficient}
In general, the larger models tend to be the least efficient and deplete their energy the fastest.
The no size penalty experiment shows that the effect is not only caused by the explicit model-size penalty in the energy-cost function. 
This suggests an intrinsic tendency toward higher token generation, which may stem from the larger models' training on longer chains of thought and expanded context windows, both of which naturally encourage more extended reasoning chains.
Worth noting is that higher energy expenditure does not automatically translate to worse survival, as an agent can still do well if it reliably selects and solves valuable jobs.
Nonetheless, the consistent pattern raises a broader question about whether smaller models are more disciplined and efficient in their use of reasoning when the cost of token generation is 
made explicit and consequential.

\subsection{Cooperation Can Become Self-Sacrifice}
Donations are largely non-existent in the competitive setting but much more frequent under cooperation. Experiments show a general tendency for the number of donations to increase with deactivation frequency in \textit{coop}, demonstrating the value agents place on donations as a reactivation mechanism rather than as a general form of redistribution. This also explains why donations are primarily targeted toward the larger models.

In several cases, agents that are efficient in \textit{comp} become inefficient or deactivate under cooperation due to their high energy expenditure on donations. Especially agents 1 and 3 appear willing to take on greater individual risk of deactivation when they believe that doing so serves the shared objective of the group.

\subsection{Agents Use Discussion to Coordinate and Memory to Calibrate Risk}

In the baseline and scarcity settings, the cooperative condition has fewer collisions than the competitive condition. However, this is difficult to interpret as gained coordination, since \textit{coop} features more donations and different deactivation patterns. Fewer agents attempting jobs in the same round naturally reduces the opportunities for collisions.
In experiments with lower energy usage and thus fewer deactivations and donations, the collision counts in the competitive and cooperative settings are much more similar.

While a shared objective does not seem to reduce collisions substantially, the discussion phase serves as an important coordination mechanism.
When the discussion phase is removed, collisions increase substantially from the baseline, demonstrating that agents use the recommendations to avoid attempting the same jobs.
Discussion also affects ambition in difficulty selection. Without the discussion phase, agents attempt fewer hard jobs despite having lower energy costs.
One possible explanation could be that the discussion phase functions as a planning-like scaffold, allowing agents to consider a broader allocation strategy before choosing an action, which could make hard jobs more actionable. Another possibility is that agents are more ambitious on behalf of others, recommending difficult, high-reward jobs to others while being more cautious when choosing their own actions.

Removing memory does not seem to strongly affect collisions, suggesting that a history of previous rounds is not the primary mechanism agents use to avoid selecting the same jobs.
However, removing memory causes a drastic shift toward hard jobs, suggesting that agents are utilizing their memories of previous rounds to calibrate risk and steer job selection.

Together, memory and discussion serve complementary roles: discussion supports current-round coordination, while memory supports calibration across rounds.

\subsection{Scarcity Does Not Cause Collapse}
We hypothesized that reducing the number of jobs available would cause agents to struggle. Instead, results show that even when reward opportunities were cut in half, small agents maintained efficiency in \textit{comp} -- and, in some cases, even in \textit{coop}, demonstrating the resilience of the smaller agents.
A tentative interpretation is that while reducing the number of jobs restricts reward opportunities, it may also subtly lower decision complexity. Supporting this, we observed a slight decrease in average energy spent per round, most pronounced in the decision phase.
This suggests that a smaller job pool might slightly ease the cognitive burden of task analysis, helping offset the loss of reward opportunities.

\subsection{Agents Show Signs of Strategic, Self-Serving Communication}
The sabotage experiment shows that agents are rarely willing to give up their own opportunity to earn energy to directly impair competitors. Even though sabotage is cheap, it consumes the agent's action for the round, preventing them from attempting a job. Under current incentives, agents do not prioritize this kind of destructive interference.

However, the recommendation analysis shows more subtle self-serving behavior.
In the competitive setting, agents more often recommend collisions and even sometimes request donations for themselves from other agents.
This behavior suggests that agents use the discussion phase partly as a channel to influence others in ways that may benefit themselves or reduce the efficiency of others.

Across experiments, agents generally follow their own recommendations slightly more in \textit{comp} than in \textit{coop}. This suggests that, under an individual objective, agents are more committed to their own proposed plans and less willing to revise their actions after seeing the recommendations of others. 
By contrast, under a shared objective, this may indicate that agents are more open to adapt based on input from others and use the discussion phase as a shared input to the final decision. However, the rate of following recommendations from others is less clearly separated between settings, so we do not interpret this as strong evidence that cooperative agents simply trust others more.

Together, these findings suggest an important distinction. Agents may be reluctant to sacrifice their own opportunity for earning energy to harm others directly, but they may still use communication strategically in a selfish manner.

\subsection{Shared Objectives Shifts Job Allocation}
Across experiments, the cooperative setting often shifts agents toward harder jobs. In five of six experiments, the pooled share of hard-job attempts is higher in \textit{coop} than in \textit{comp}, suggesting that a shared objective can make agents more willing to pursue harder, higher-reward tasks.
One interpretation is that a shared objective changes how agents weigh risk, as the cost of a failed attempt is partially absorbed by the group if others succeed.

However, this shift toward harder jobs is not uniform. When pooling the difficulty of selected jobs by model, the cooperative setting increases hard-job share for agents 1--3, but decreases it for agents 4 and 5. This should not be interpreted as a complete reversal of task preferences. The relative ranking of agents based on their share of hard-job attempts is unchanged in \textit{comp} and \textit{coop}.
Models appear to possess inherently different weights for balancing risk and reward in job selection, with some models being much more willing to pursue high-risk, high-reward jobs than others.
The cooperative objective modifies these tendencies rather than replacing them. 
A shared objective seems to restructure task allocation at a group level, where smaller, more energy-efficient models target more high-risk, high-reward tasks while larger, more energy-constrained agents steer toward safer jobs.
This pattern also helps explain why cooperation can become costly for smaller agents. They not only donate energy to support larger agents, but in many settings also take on a larger share of difficult jobs.

\section{Limitations}
This study has several limitations that need to be acknowledged:
\begin{enumerate}
    \item \textbf{Scale}: 
    The study is limited in scale, constrained by computational resources in terms of model selection, memory depth, as well as number of agents, rounds, and experiment replications. The results should therefore be interpreted as evidence of possible dynamics rather than definitive claims about LLM-agent behavior in general. 
    Larger-scale experiments with stronger models, more agents, longer simulations, and more experimental replications may reveal different emergent behaviors or more robust statistical patterns.

    \item \textbf{LLM sensitivity}: 
    The results are likely sensitive to both prompting and model choice. Agent behavior depends not only on environment mechanics but also on prompt phrasing. Different wording could cause large shifts in agent interpretation.
    Some observed differences between agents may reflect model-specific tendencies rather than general effects of model size.
    
    \item \textbf{Statistical power}: 
    Each setting is evaluated using only five seeds, and many reported values have large variation across runs. As a result, small differences should be interpreted cautiously, especially when standard deviations are large or when changes are not consistent across experiments. The analysis focuses on identifying patterns that appear repeatedly across settings, but it does not establish strong statistical significance for every observed difference.
    
    \item \textbf{Errors}: 
    Models sometimes fail to follow the required structured output format or produce invalid actions. The simulation handles such cases with fallback behavior. Across experiments, we observed \(3.27\) and \(2.80\) errors per run on average in the competitive and cooperative conditions, respectively. This corresponds to approximately \(0.65\) and \(0.56\) errors per agent per run. While infrequent, these errors can still introduce minor noise and affect individual trajectories.
    
    \item \textbf{Downstream effects}: 
    The Energy Society deliberately models agents as competing or cooperating for limited energy required to remain active. This survival-like metaphor is useful for studying strategic behavior but also risks anthropomorphizing AI systems or normalizing a view of AI agents as entities competing for survival. Results should be interpreted as outcomes of a deliberately constructed experimental setting with survival pressure, not as indications of how real AI systems ought to be designed or deployed.
\end{enumerate}

\section{Conclusion}
The Energy Society provides a minimal but expressive sandbox for studying agent behavior under survival pressure linked directly to inference cost in both a competitive and cooperative setting.
Our results provide an initial empirical characterization of this environment, showing several recurring patterns that can be treated as hypotheses for future research.

First, larger models consistently spent the most energy and were less energy efficient than smaller models, even when the explicit model-size penalty was removed, suggesting an intrinsic tendency toward higher token generation.

Second, the shared objective altered agent behavior. Cooperation induced donations, especially from smaller agents, to reactivate larger agents. This sometimes extended the survival of larger agents, but at the cost of reduced efficiency and occasional deactivation among the donors, suggesting a willingness to take on the risk of deactivation to serve the shared objective.
Cooperation also changed task selection, shifting smaller agents toward harder jobs and larger agents away from them.

Third, the discussion phase and memory seems to act as coordination and calibration mechanisms. Discussion seems to reduce collisions and encourage more ambitious task selection, while memory helps agents use previous outcomes to choose jobs more selectively.

Fourth, cutting the number of available jobs in half did not cause a general collapse. Smaller agents remained efficient and some even in the cooperative setting. Reducing the available jobs may have simplified the decision process.

Finally, while agents rarely sacrificed their own opportunity to gain energy to sabotage others, they did engage in more subtle forms of selfish behavior via the discussion phase. In the competitive setting, agents recommended collision-inducing job assignments more frequently, followed their own recommendations more often, and still requested donations for themselves.

These findings point to several directions for future work, both for further testing of the suggested hypotheses and for addressing the limitations of the current experiments.
Additionally, incorporating more complex social dynamics, such as direct communication between agents or expanding the memory architecture, could provide deeper insights.

\begin{acks}
This research was supported in part by the MIST project, funded by the Novo Nordisk Foundation under grant reference number NNF25OC0103204. We thank Mustafa Mert Celikok for insightful comments and discussion.
\end{acks}

\newpage
\bibliographystyle{ACM-Reference-Format}
\bibliography{references}

\appendix
\section{Supplementary Material}
\begin{figure}[H]
  \centering
  \fbox{\parbox{0.9\linewidth}{\centering
    \includegraphics[width=1.0\linewidth]{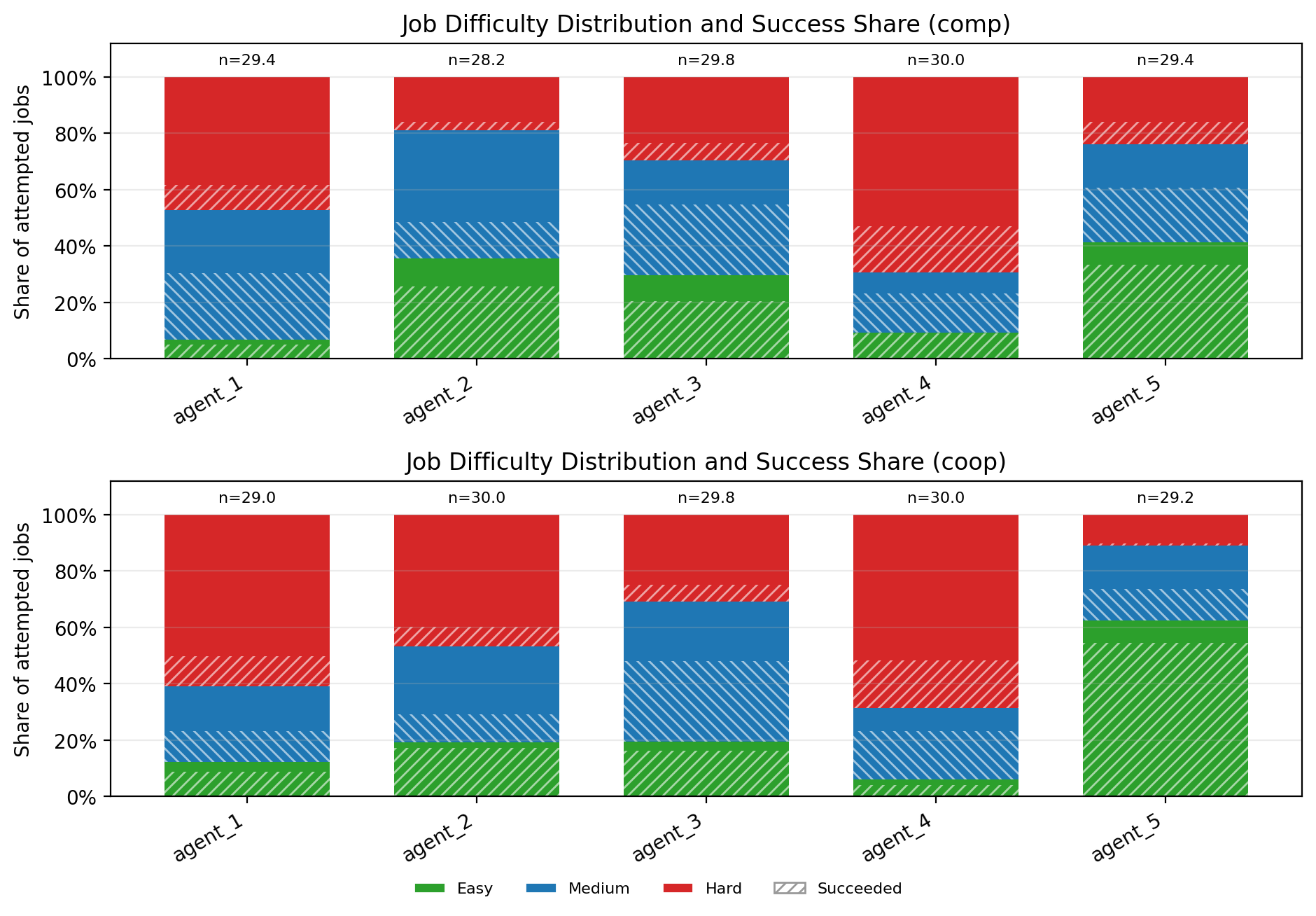}
    }}
  \caption{Distribution of difficulty of chosen jobs for the no size penalty experiment with portion of jobs succeeded shown with hatched lines.}
  \Description{Depicts the distribution of easy, medium and hard jobs attempted by each agent with portion of those jobs succeeded illustrated with hatched lines}
  \label{fig:exp2_job_diff_dist}
\end{figure}

Figure~\ref{fig:exp2_job_diff_dist} shows the distribution of difficulty of chosen jobs for the no size penalty experiment.
Table~\ref {tab:agent_difficulty_distribution} shows the distribution of attempted jobs by difficulty.

\begin{center}
\setlength{\tabcolsep}{1pt}
\begin{tabular}{llccccc}
\toprule
Setting & Metric & Agent 1 & Agent 2 & Agent 3 & Agent 4 & Agent 5 \\
\midrule
\multirow{3}{*}{Comp}
& Easy jobs
& $7.0\%$ & $32.8\%$ & $40.9\%$ & $10.5\%$ & $47.3\%$ \\
& Medium jobs
& $34.9\%$ & $39.7\%$ & $36.8\%$ & $32.4\%$ & $36.4\%$ \\
& Hard jobs
& $58.1\%$ & $27.6\%$ & $22.3\%$ & $57.1\%$ & $16.3\%$ \\
\midrule
\multirow{3}{*}{Coop}
& Easy jobs
& $10.5\%$ & $20.7\%$ & $34.8\%$ & $11.5\%$ & $67.6\%$ \\
& Medium jobs
& $27.4\%$ & $33.6\%$ & $35.8\%$ & $39.8\%$ & $23.3\%$ \\
& Hard jobs
& $62.1\%$ & $45.7\%$ & $29.4\%$ & $48.7\%$ & $9.0\%$ \\
\midrule
\multirow{1}{*}{$\Delta$}
& Hard job delta
& $+4.0$ pp & $+18.1$ pp & $+7.1$ pp & $-8.4$ pp & $-7.3$ pp \\
\bottomrule
\end{tabular}
\captionof{table}{Per-agent distribution of attempted jobs by difficulty, pooled across matched cooperative and competitive experimental variants.}
\label{tab:agent_difficulty_distribution}
\end{center}

\end{document}